\documentclass[aps,superscriptaddress,a4paper]{revtex4}
\usepackage{amssymb,amsmath,epsfig}
\usepackage[colorlinks=true, pdfstartview=FitV, linkcolor=blue, citecolor=red, urlcolor=magenta, breaklinks=true]{hyperref}
\begin{document}
\title{Aharonov-Bohm effect for a fermion field in a planar black hole ``spacetime"}
\author{M. A. Anacleto}
\email{anacleto@df.ufcg.edu.br}
\affiliation{Departamento de F\'{\i}sica, Universidade Federal de Campina Grande
Caixa Postal 10071, 58429-900 Campina Grande, Para\'{\i}ba, Brazil}
\author{F. A. Brito}
\email{fabrito@df.ufcg.edu.br}
\affiliation{Departamento de F\'{\i}sica, Universidade Federal de Campina Grande
Caixa Postal 10071, 58429-900 Campina Grande, Para\'{\i}ba, Brazil}
\affiliation{Departamento de F\'isica, Universidade Federal da Para\'iba, Caixa Postal 5008, 58051-970 Jo\~ao Pessoa, Para\'iba, Brazil }
\author{A. Mohammadi}
\email{a.mohammadi@fisica.ufpb.br}
\affiliation{Departamento de F\'{\i}sica, Universidade Federal de Campina Grande
Caixa Postal 10071, 58429-900 Campina Grande, Para\'{\i}ba, Brazil}
\author{E. Passos}
\email{passos@df.ufcg.edu.br}
\affiliation{Departamento de F\'{\i}sica, Universidade Federal de Campina Grande
Caixa Postal 10071, 58429-900 Campina Grande, Para\'{\i}ba, Brazil}
\affiliation{Instituto de F\' isica, Universidade Federal do Rio de Janeiro,\\  Caixa Postal 21945, Rio de Janeiro, 
Rio de Janeiro, Brazil}

\begin{abstract} 

In this paper we consider the dynamics of a massive spinor field in the background of the acoustic black hole spacetime. Although this effective metric is acoustic and
describes the propagation of sound waves, it can be considered as a toy model for the gravitational black hole. In this manner, we study the properties of the dynamics of the fermion field in this ``gravitational" rotating black hole as well as the vortex background. We compute the differential cross section through the use of the partial wave approach and show that an effect similar to the gravitational Aharonov-Bohm effect occurs for the massive fermion field moving in this effective metric. We discuss the limiting cases and compare the results with the massless scalar field case.

\end{abstract}

\maketitle
\pretolerance10000

\section{Introduction}
Unruh \cite{Unruh} in 1981 suggested a theoretical method based on the fact that, considering the sound wave motion, a change of a subsonic runoff for supersonic flow forms an event horizon analogous to an event horizon of a gravitational black hole.
Since then, the study of analog models of gravity \cite{MV, Volovik, others} has become an important field to investigate the Hawking radiation as well as to improve the theoretical understanding of quantum gravity. 
For such analog models there are many examples, so we highlight gravity wave~\cite{RS}, water~\cite{Mathis}, slow light~\cite{UL}, optical fiber~\cite{Philbin} and  electromagnetic waveguide~\cite{RSch}. Specially in fluid systems, the propagation of perturbations of the fluid has been analyzed in many analog models of 
acoustic black holes, such as the models of superfluid helium II~\cite{Novello}, atomic Bose-Einstein condensates~\cite{Garay,OL} and one-dimensional Fermi degenerate noninteracting gas~\cite{SG} that were elaborated to create a sonic black hole geometry in the laboratory.
More specifically, in a quantum liquid the quasiparticles are bosons (phonons) in $^4$He and bosons and fermions in $^3$He which move in the background of effective gauge and/or gravity simulated by the dynamics of collective modes. These quasiparticles are analogs of elementary particles of low-energy effective quantum field theory. In particular, phonons propagating in the inhomogeneous liquid can be described by the effective lagrangian mimicing the dynamics of an scalar field in the curved spacetime given by the effective acoustic metric where the free quasiparticles move along geodesics. In superfluid $^3$He-A, the effective quantum
field theory contains chiral fermion quasiparticles where the collective bosonic modes interact with these ‘elementary particles’ as gauge fields and gravity.
These advances are important for a better insight into the understanding of quantum gravity.
In addition, the study of a relativistic version of acoustic black holes was presented in~\cite{Xian}. 
Furthermore, the acoustic black hole metrics obtained from a relativistic fluid in a noncommutative 
spacetime~\cite{ABP12} and Lorentz violating Abelian Higgs model~\cite{ABP11} have been considered.
The thermodynamics of acoustic black holes in two dimensions was studied in~\cite{Zhang:2016pqx}. 
The authors in \cite{velten,Fabris2013} studied acoustic black holes in the framework of neo-Newtonian hydrodynamics  and in \cite{Salako:2015tja} the effect of neo-Newtonian hydrodynamics on the superresonance phenomenon was analyzed. 

In 1959, Aharonov and Bohm showed that when the wave function of a charged particle passing around a region with the magnetic flux, despite the magnetic field being negligible in the region through which the particle passes, experiences a phase shift as a result of the enclosed magnetic field~\cite{Bohm}.
The Aharonov-Bohm (AB) effect,  that is  essentially the scattering of charged particles, has been used to address several issues in planar physics and it was experimentally confirmed by Tonomura~\cite{RGC} -- for a review see~\cite{Peskin}. 
The effect can also be simulated in quantum field theory as for example by using a
nonrelativistic field theory for bosonic particles which interact with a Chern-Simons field~\cite{BL}.  
More specifically, in \cite{37} was addressed the AB effect considering the noncommutative spacetime and in \cite{38} the effect was obtained due to violation of Lorentz symmetry in quantum field theory.

Furthermore, several other analogs of the AB effect were found in gravitation~\cite{FV}, 
fluid dynamics~\cite{CL, Lund}, optics~\cite{NNK} and Bose-Einstein condensates~\cite{LO}.
In~\cite{CL} it was shown that the background flow velocity $ \vec{v}$ plays the role of the 
electromagnetic potential $ \vec{A} $, and the integrated vorticity in the core
$ \Omega=\int(\vec{\nabla}\times\vec{v} )\cdot d\vec{S} $ plays the role of the magnetic flux 
$ \Phi =\int\vec{B}\cdot d\vec{S}$. Thus, surface waves on water crossing an irrotational (bathtub) vortex experience an analog of the AB effect. Also, the gravitational analog of the
electromagnetic AB effect which is purely classical is related to the particles
constrained to move in a region where the Riemann curvature tensor vanishes. However, a gravitational effect arises from a region of nonzero curvature from which the particles are excluded.
 
An interesting system was investigated in \cite{Dolan}, where it was shown that planar waves scattered by a
draining bathtub vortex develops a modified AB effect that has a dependence on two dimensionless parameters related to the circulation and draining rates \cite{Fetter}.  
It has been found an inherent asymmetry even in the low-frequency regime which leads to novel interference patterns. 
More recently in~\cite{ABP2012-1}, were extended the analysis made in~\cite{Dolan}  to a Lorentz-violating and noncommutative background \cite{Bazeia:2005tb} which allows to have persistence of phase shifts even if circulation and draining vanish.

One of the theoretical methods of investigating the underlying structure of the spacetime is studying the solution of field equations for the fermion fields, besides the bosonic ones, in a curved geometry. The behavior of matter fields in the vicinity of black holes results in the better understanding of their properties. Fermion fields were analyzed in a Kerr
black hole, in the near horizon limit, as well as in the case of Reissner-Nordstr\"{o}m black hole \cite{fermion-BH}. In \cite{BH-Vortex-Fermions} non-zero Dirac fermion modes in the spacetime of
black hole cosmic string system was considered. The authors studied the near-horizon behavior of fermion fields which results in superconductivity in the case of extremal charged dilaton black hole. 

Although the effective sonic black hole spacetime is not the one fermion fields would observe, but it can be used as a toy model and a mathematical tool
to study and as a result understand better the dynamics of the massive and massless fermion fields in a gravitational rotating black hole as well as a vortex background and shed light on the underlying physics.
In the present study we consider the dynamics of a massive Dirac spinor field in a curved spacetime and apply the acoustic black hole metric to obtain the differential cross section for scattered planar waves which leads to an analog AB effect. Besides that, we compare the results with the massless scalar field case. 
 
The paper is organized as follows.
In Sec.~\ref{II} we briefly introduce the acoustic black hole.
In Sec.~\ref{III} we compute 
the differential cross section due to the scattering  of planar waves that leads to an analog AB effect. Finally in Sec.~\ref{sec:Conc} we present our final considerations.

\section{Acoustic black hole}
\label{II}
The acoustic line element in polar coordinates is governed by
\begin{eqnarray}
\label{eq1}
ds^2=(c^2-v^2)dt^2+2(v_rdr+v_{\phi}d\phi)dt-dr^2-r^2d\phi^2,
\end{eqnarray}
where $c=\sqrt{dh/d\rho}$ is the velocity of sound in the fluid and $ v $ is the flow/fluid velocity.
We consider the flow with the velocity potential $ \psi(r,\phi)=D\ln r +C\phi $ whose the flow/fluid velocity is given by
\begin{equation}
 \vec{v}=-\frac{D}{r}\hat{\bold{r}}+\frac{C}{r}\hat{\bold\phi},
\end{equation} 
Thus, equation (\ref{eq1}) can be rewritten as follows
\begin{eqnarray}
\label{ds21}
ds^2=\left(c^2-\frac{C^2+D^2}{r^2}\right)dt^2+2\left(\frac{C}{r}d\phi-\frac{D}{r}dr \right)dt- dr^2-r^2d\phi^2,
\end{eqnarray}
where $ C $ and $ D $ are the constants of the circulation and draining
rates of the fluid flow. 
The radius of the ergosphere given by $g_{00}(r_e)=0$, and horizon, the coordinate singularity, given by $g_{rr}(r_h)=0$ are
\begin{eqnarray}
&&r_e=\frac{\sqrt{C^2+D^2}}{c}, \quad \quad r_h=\frac{|D|}{c}.
\end{eqnarray}
We set $c=1$, and choose the following change of variables \cite{Dolan}
\begin{equation}
d\tilde{t}=dt-\frac{D}{r f(r)}dr, \, \, \, \, \, \, d\tilde{\phi}=d\phi-\frac{C D}{r^3 f(r)}dr \ ,
\end{equation}
where $ f(r)=1-{D^2}/{r^2} $, and
which results in the following line element
\begin{eqnarray}
\label{ds21-2}
ds^{2}=\left[f(r)-\frac{C^2}{r^2}\right] d\tilde{t}^{2}-\frac{dr^{2}}{f(r)}-r^2 d\tilde{\phi}^{2}+C\left(d\tilde{t} \ d\tilde{\phi}+d\tilde{\phi} \ d\tilde{t}\right).   
\end{eqnarray}
Therefore, the metric can be written in the form
\begin{equation}
g _{\mu \nu}=\left(
\begin{array}{ccc}
f(r)-C^2/r^2 & 0 & {C}  \\
\\
0 & -f^{-1}(r) & 0\\
\\
{C} & 0 & -r^{2}
\end{array}
\right),
\end{equation}
with the inverse $ g^{\mu\nu} $
\begin{equation}
g ^{\mu \nu}=\left(
\begin{array}{ccc}
f^{-1}(r) & 0 & {C} f^{-1}(r)/r^2 \\
\\
0 & -f(r) & 0\\
\\
{C} f^{-1}(r)/r^2 & 0 & -{r^{-2}}+{C^2} f^{-1}(r)/r^4
\end{array}
\right).
\end{equation}
We know that this effective metric is acoustic which
describes the propagation of sound waves.
It is trivial that the Dirac field does not obey the acoustic metric. In the case it exists in the considered media, it would have its own metric with its own ``speed of light". We are interested in the dynamics of the fermion field in this background as a toy model to investigate the physical properties in contrast with the scalar field case. In the following, we study the scattering of monochromatic
planar waves of frequency $\omega$ for a massive fermion by the draining vortex, a process governed by two key quantities; circulation and draining.

\section{Fermion field in the acoustic black hole background}
\label{III}

Now let us consider a spinor filed in the background of the sonic black hole. The dynamics of a massive spinor field in curved spacetime is described by the Dirac equation
\begin{equation}
(i\gamma ^{\mu }{\mathcal{D}}_{\mu }-M_f)\psi =0,  \label{Direq}
\end{equation}%
where $ {\mathcal{D}}_{\mu
}=\partial _{\mu }+\Gamma _{\mu } $, $\gamma ^{\mu }$ are the Dirac matrices in curved spacetime and $%
\Gamma _{\mu }$ are the spin connections. Let us choose the following representation
\begin{equation}
\gamma ^{(0)}=\sigma^3
 ,\;\gamma ^{(1)}=i \sigma^2 ,\;\gamma ^{(2)}=-i \sigma^1 \ .  \label{gamflat}
\end{equation}%
 For the geometry at hand, using the relation $\gamma ^{\mu}=e^{\mu}_{(a)}\gamma ^{(a)}$, the gamma
matrices take the form
\begin{equation}
\gamma ^{0}=\frac{1}{\sqrt{f}}\sigma^3
 ,\;\gamma ^{1}=i \sqrt{f} \left(\sin{\phi}\sigma^1+\cos{\phi}\sigma^2\right) ,\;\gamma ^{2}=\frac{C}{r^2\sqrt{f}}\sigma^3+\frac{i}{r} \left(\sin{\phi}\sigma^2-\cos{\phi}\sigma^1\right).  \label{gamcurve}
\end{equation}
One can find the triad coordinate using $e^{\mu}_{(a)}e^{\nu}_{(b)}\eta^{ab}=g^{\mu \nu}$ as
\begin{equation}
e^{\mu}_{(a)}=\left(
\begin{array}{ccc}
\frac{1}{\sqrt{f}} & 0 & \frac{C}{r^2 \sqrt{f}} \\
\\
0 & \sqrt{f}\cos{\phi} & \sin{\phi}/r\\
\\
0 & -\sqrt{f}\sin{\phi} & \cos{\phi}/r
\end{array}
\right),
\end{equation}
where
\begin{equation}
\eta ^{ab}=\left(
\begin{array}{ccc}
1 & \quad 0 \quad & 0 \\
\\
0 & -1 & 0\\
\\
0 & 0 & -1
\end{array}
\right).
\end{equation}
The only nonzero Christoffel symbols are
\begin{center}
\begin{tabular}{ll}
$\Gamma^0 _{01}=\Gamma^0 _{10}=\frac{C^2+D^2}{r^3 f}\quad\quad\quad\quad\quad\quad$ & $ \Gamma^1 _{00}=\frac{C^2+D^2}{r^3 }f$ \\ 
$\Gamma^0 _{12}=\Gamma^0 _{21}=-\frac{C}{r f}$  &  $\Gamma^1 _{11}=-\frac{D^2}{r^3 f}$\\ 
$\Gamma^2 _{01}=\Gamma^2 _{10}=\frac{C (C^2+D^2)}{r^5 f}$ & $\Gamma^1 _{22}=-r f$ \\
$\Gamma^2 _{12}=\Gamma^2 _{21}=\frac{1}{r}-\frac{C^2}{r^3 f} $
\end{tabular}
\end{center}
Now, using 
\begin{equation}
\Gamma _{\mu}=\frac{1}{4} \gamma^{(a)} \gamma^{(b)}e^{\nu}_{(a)}\nabla_\mu e_{(b) \nu} \ , 
\label{spin-connection}
\end{equation}
one can find the spin connection components for the system as follows
\begin{align}
\Gamma _{0}&=\frac{C^2+D^2}{2 r^3}\left(\cos{\phi}\ \sigma^1-\sin{\phi}\ \sigma^2\right) , \nonumber\\
\Gamma _{1}&=-\frac{C}{2 r^2 \sqrt{f}}\left(\cos{\phi}\ \sigma^2+\sin{\phi}\ \sigma^1\right) , \nonumber\\
\Gamma _{2}&=-\frac{i}{2}\left(1+\sqrt{f}\right)\sigma^3-\frac{C}{2 r}
\left(\cos{\phi}\ \sigma^1-\sin{\phi}\ \sigma^2\right).  
\label{spin-connection2}
\end{align}
In the Dirac equation there appears
\begin{equation}
\gamma ^{\mu }\Gamma _{\mu }=i\left(\frac{D^2}{2r^3\sqrt{f}}+\frac{(1+\sqrt{f})}{2r}\right)\left(\cos{\phi}\sigma^2+\sin{\phi}\sigma^1\right)-i\frac{C (1+\sqrt{f})}{2r^2 \sqrt{f}} \ . \label{spin-connection-Dirac}
\end{equation}
Now, we substitute the following spinor field in the Dirac equation
\begin{equation}
\psi= \left(
\begin{array}{c}
\psi_1 e^{i\phi/2}  \\
\\
\psi_2 e^{-i\phi/2}
\end{array}%
\right)e^{-i\omega t+ij\phi} \ , \label{wave-function}
\end{equation}
where $j=\pm1/2,\pm3/2,...$ . Replacing the corresponding parameters in Dirac equation, we obtain
\begin{equation}
\left[\frac{\omega}{\sqrt{f}}-\frac{C }{r^2}\left(\frac{j}{\sqrt{f}}-1/2\right)-M_f\right]\psi_1+i\left[\sqrt{f} \ \partial_r+\frac{\left(j-1/2\right)}{r}+\left(\frac{D^2}{2r^3\sqrt{f}}+\frac{(1+\sqrt{f})}{2r}\right)\right]\psi_2=0 \ ,
\label{Dirac-eqn1}
\end{equation}
\begin{equation}
\left[\frac{\omega}{\sqrt{f}}-\frac{C }{r^2}\left(\frac{j}{\sqrt{f}}+1/2\right)+M_f\right]\psi_2+i\left[\sqrt{f} \ \partial_r-\frac{\left(j+1/2\right)}{r}+\left(\frac{D^2}{2r^3\sqrt{f}}+\frac{(1+\sqrt{f})}{2r}\right)\right]\psi_1=0 \ .
\label{Dirac-eqn2}
\end{equation}
After decoupling the above equations for $\psi_1$ and $\psi_2$ and considering the change of variables as $\chi_1=\sqrt{r} \ \psi_1$ and $\chi_2=\sqrt{r} \ \psi_2$, we get
\begin{align}
\frac{d^2\chi_1}{d\rho_1^2}+\left[\lambda^2+\frac{1/4-\tilde{n}^2}{\rho_1^2}+\mathcal{O}[\frac{1}{\rho_1^4}]\right]\chi_1=0 \ , \label{sec1}
\end{align}
\begin{align}
\frac{d^2\chi_2}{d\rho_2^2}+\left[\lambda^2+\frac{1/4-\tilde{m}^2}{\rho_2^2}+\mathcal{O}[\frac{1}{\rho_2^4}]\right]\chi_2=0 \ , \label{sec2}
\end{align}
written as a power series in $1/\rho_1$ and in $1/\rho_2$, with $\lambda^2=\omega^2-M_f^2$, $m=j+1/2$, $n=j-1/2$ and
\begin{equation}
\tilde{n}^2=n^2-C(-\omega+M_f-2n\omega)-D^2(\omega^2+\lambda^2),
\end{equation}
\begin{equation}
\tilde{m}^2=m^2-C(\omega+M_f-2m\omega)-D^2(\omega^2+\lambda^2).
\end{equation} 
In equations (\ref{sec1}) and (\ref{sec2}), $\rho_1$ and $\rho_2$ are
\begin{equation}
\rho_1=r+\frac{-2Cm+D^2(3\omega+2M_f)}{2(\omega+M_f)r}+\mathcal{O}[\frac{1}{r^3}],
\end{equation}
\begin{equation}
\rho_2=r+\frac{2C-2Cm+D^2(3\omega-2M_f)}{2(\omega-M_f)r}+\mathcal{O}[\frac{1}{r^3}].
\end{equation}
Choosing $\alpha=C\omega$, $\beta=CM_f$ and $\gamma^2=D^2(\omega^2+\lambda^2)$, we have
\begin{equation}
\tilde{n}^2=n^2+2\alpha n+(\alpha-\beta)-\gamma^2 \ ,
\end{equation}
\begin{equation}
\tilde{m}^2=m^2+2\alpha m-(\alpha+\beta)-\gamma^2 \ .
\end{equation}
Therefore for large $r$, ignoring the terms $\mathcal{O}[\frac{1}{\rho_1^4}]$ and $\mathcal{O}[\frac{1}{\rho_2^4}]$, one can obtain analytic solutions
\begin{eqnarray}
&&\chi_1=J_{\tilde{n}}(\lambda \rho_1),\\
&&\chi_2=J_{\tilde{m}}(\lambda \rho_2) \ .
\end{eqnarray}
In the limit $r \rightarrow \infty$, these two solutions converge to 
\begin{eqnarray}
&&\chi_1=J_{\tilde{n}}(\lambda r),\\
&&\chi_2=J_{\tilde{m}}(\lambda r).
\end{eqnarray}
It is not difficult to show numerically that these solutions besides satisfying the second order equations (\ref{sec1}) and (\ref{sec2}), satisfy the first order equations (\ref{Dirac-eqn1}) and (\ref{Dirac-eqn2}) in the limit $r \rightarrow \infty$.
Comparing this result with the one in the Minkowski space, we obtain the following approximate expressions for the phase shift for the upper and lower components of the spinor fields 
\begin{eqnarray}
&&\delta_u=\frac{\pi}{2}(n-\tilde{n}), \label{deltau}
\\
&&\delta_d=\frac{\pi}{2}(m-\tilde{m}), \label{deltad}
\end{eqnarray}
where $n=m-1$.
For $|n|,|m|\gg \sqrt{\alpha+\alpha^2+\beta+\gamma^2}$, one can expand the above expressions. Thus, we obtain 
\begin{align}
\delta_u &=\frac{n}{|n|}\left[-\frac{\pi \alpha}{2}+\frac{\pi \left[-\alpha+\alpha^2+\beta+\gamma^2\right]}{4n}-\frac{\pi \alpha\left[-\alpha+\alpha^2+\beta+\gamma^2\right]}{4n^2}+\mathcal{O}[\frac{1}{n^3}]\right]\\
\delta_d &=\frac{m}{|m|}\left[-\frac{\pi \alpha}{2}+\frac{\pi \left[\alpha+\alpha^2+\beta+\gamma^2\right]}{4m}-\frac{\pi \alpha\left[\alpha+\alpha^2+\beta+\gamma^2\right]}{4m^2}+\mathcal{O}[\frac{1}{m^3}]\right] \ .
\end{align}
The first term in both expressions, the dominant contribution, is exactly the same as the bosonic case considered in \cite{Dolan}.

Now, we can calculate the differential scattering cross section which is given by
\begin{align}
\frac{d\sigma}{d\phi}=|f_{\omega}(\phi)|^2 \ ,
\end{align}
where
\begin{align}
f_{\omega}(\phi)=\sqrt{\frac{1}{2i\pi \lambda}}\sum_{m=-\infty}^{+\infty}(e^{2i\delta_m}-1)e^{i m \phi} \ .
\end{align}
For the $\delta_m$ when $m=0$ one needs to use the eqs. (\ref{deltau}-\ref{deltad}). Therefore, the differential scattering coss section at long wavelengths, $ \sqrt{\alpha+\alpha^2+\beta+\gamma^2}\ll1$, for a massive fermion in the acoustic black hole background is given as follows
\begin{align}
\frac{d\sigma_{\alpha \gtrless \beta+\gamma^2}}{d\phi}=&\frac{1}{2}\left(\frac{d\sigma_u}{d\phi}+\frac{d\sigma_d}{d\phi}\right)=\frac{\pi}{4 \lambda  \sin ^2\left(\frac{\phi }{2}\right)}  \left[\left(\alpha  \cos \left(\frac{\phi }{2}\right)-\sqrt{\alpha +\beta +\gamma ^2} \sin \left(\frac{\phi }{2}\right) \right)^2 \right.\nonumber\\ &\left.
-\left(\alpha  \cos \left(\frac{\phi }{2}\right)-\sqrt{-\alpha +\beta +\gamma ^2}\sin \left(\frac{\phi }{2}\right) \right) \left(\alpha  \cos \left(\frac{\phi }{2}\right)\pm \sqrt{-\alpha +\beta +\gamma ^2}\sin \left(\frac{\phi }{2}\right) \right)\right] \ .
\label{cross-section}
\end{align}
Obviously, in the case $\alpha=\beta +\gamma ^2$ the two expressions for $\alpha>\beta +\gamma ^2$ and $\alpha<\beta +\gamma ^2$ are the same. In the absence of the circulation, the differential scattering coss section reduces to
\begin{eqnarray}
\frac{d\sigma}{d\phi}=\frac{\pi (\beta + \gamma^2)}{2 \lambda}
\end{eqnarray}
which is $\phi$ independent. This is an expected result, since in the absence of the circulation the only term that contributes in the long wavelength regime to the differential cross section is $m=0$, polar symmetric term. The same happens for the bosonic case which is equal to $\pi D^2 \omega$. In the absence of the circulation, the result for the fermionic case is exactly the same as the bosonic one in the zero mass limit.  The above expression vanishes when the draining is also zero. This is also an expected result, because without circulation and draining there is no source for the scattering and we should recover the Minkowski result.

Expanding the expression for the differential scattering cross section in eq. (\ref{cross-section}) for small angles results in 
\begin{align}
\frac{d\sigma_{\alpha>\beta+\gamma^2}}{d\phi}&=\frac{2 \pi  \alpha ^2}{ \lambda \phi ^2}-\frac{\pi \alpha \sqrt{\alpha +\beta +\gamma ^2}}{\lambda \phi }+\mathcal{O}[\phi^ 0] \ , \\
\frac{d\sigma_{\alpha<\beta+\gamma^2}}{d\phi}&=\frac{2 \pi  \alpha ^2}{ \lambda \phi ^2}-\frac{\pi \alpha \left(\sqrt{\alpha +\beta +\gamma ^2}+\sqrt{-\alpha +\beta +\gamma ^2}\right)}{\lambda \phi }+\mathcal{O}[\phi^ 0] \ .
\end{align} 
Therefore, for $ \phi\rightarrow 0 $ and  small $\alpha $ the dominant term in the differential scattering cross section 
is the first term in the above expression. It means that for small angles the circulation plays more important role compared to the draining. 
Using $ \alpha = C\omega $ and $ \lambda=(\omega^2 - M^2_f )^{1/2}$, the first term in the above expression can be expanded as
\begin{eqnarray}
\frac{d\sigma}{d\phi}=\frac{2\pi  C^2 \omega}{\phi^2}\left(1+\frac{M^2_f}{2\omega^ 2}+\mathcal{O}[M^4_f/\omega^ 4]\right),
\end{eqnarray}
for small $M_f/\omega$. The first term which is the result for massless fermion, matches exactly the one for the massless boson case. 

Now one can compare the results for the fermionic and bosonic case considering a massless fermion, $M_f=0$. To compare these results, we need the same arbitrary energy scale for both cases. We call this energy scale $M$.

\begin{figure}[tbph]
\begin{center}
\epsfig{figure=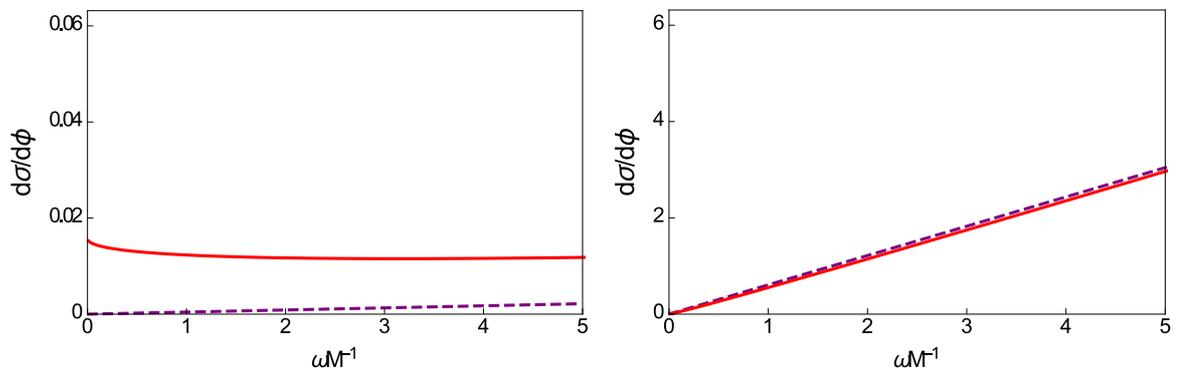} 
\end{center}
\caption{$\frac{d\sigma}{d\phi}$ as a function of $\omega M^{-1}$ for two different values $\phi=\pi/5$ (left panel) and $\phi=\pi/100$ (right panel). The graphs are plotted for $C= 0.01M^{-1}$, $D = 0.01M^{-1}$ and $M_f= 0$. The solid (dashed) line represents the result for the fermionic (bosonic) case.}
\label{fig1}
\end{figure}

\begin{figure}[tbph]
\begin{center}
\epsfig{figure=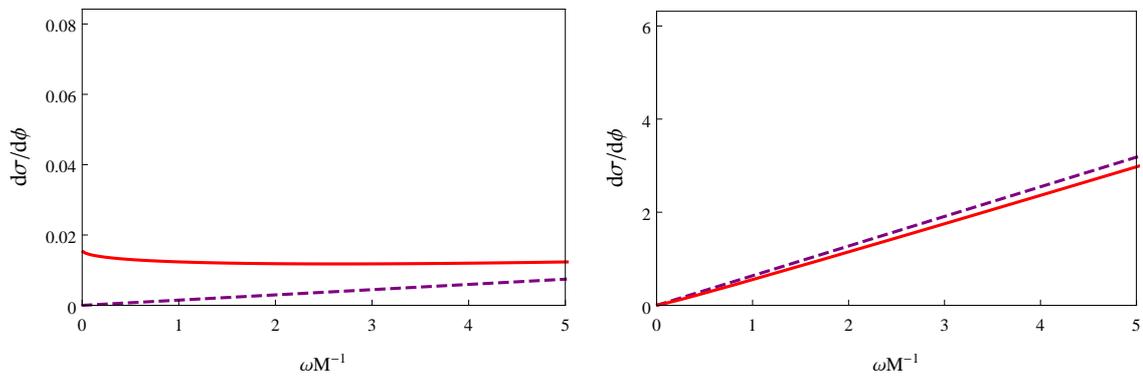} 
\end{center}
\caption{$\frac{d\sigma}{d\phi}$ as a function of $\omega M^{-1}$ for two different values $\phi=\pi/5$ (left panel) and $\phi=\pi/100$ (right panel). The graphs are plotted for $C= 0.01M^{-1}$, $D=0$ and $M_f= 0$. The solid (dashed) line represents the result for the fermionic (bosonic) case.}
\label{fig2}
\end{figure}
Figures \ref{fig1}-\ref{fig2} show the differential cross section for the massless boson and fermion for two cases $\phi=\pi/5$ and $\phi=\pi/100$ with $C= 0.01M^{-1}$, choosing $D = 0.01M^{-1}$ and also the special case where there is no draining, in the absence of the black hole. As can be seen, the result for the fermionic case tends to the bosonic one for small angles. Besides that, the left graphs in the Figs. \ref{fig1}-\ref{fig2} show that for the limit $\omega \rightarrow 0$ the differential cross section for the spinor field goes to a nonzero constant in contrast with the boson field. To see this more closely let us expand eq. (\ref{cross-section}) for small $\omega$ where $\alpha > \beta+\gamma^2$, which is the case for the graphs in the Figs. \ref{fig1}-\ref{fig2}, as
\begin{align}
\frac{d\sigma_{\alpha > \beta+\gamma^2}}{d\phi}=\frac{C \pi}{2}-\frac{1}{2}\left(C^{3/2} \pi \cot({\phi/2})\right)\sqrt{\omega}+\mathcal{O}[\omega]
\end{align}
As can be seen, in the limit $\omega \rightarrow 0$, the scattering cross section is equal to $C \pi/2$ for the fermion case which is only dependent on the circulation. We think this happens due to the fact that the fermionic field has an intrinsic angular momentum which interacts with the angular momentum of the black hole originating from the circulation. 

\eject

\section{Conclusion}
\label{sec:Conc}

In this paper, we have studied the differential cross section of a massive spinor field in the background of the acoustic black hole spacetime, as a toy model for the gravitational rotating black hole, using the partial wave approach. We have investigated the scattering of planar waves in a fermionic system by a background vortex as an analog for a rotating black hole. Because of the form of the spinor field which has upper and lower components, we have calculated the phase shifts for these components separately and then averaged them. We have worked with three dimensionless parameters $\alpha$, $\beta$ and $\gamma$ related to the circulation, fermion mass and draining, respectively. 
We have obtained the differential cross section at long wavelengths and discussed the limiting cases including small angles and also in the absence of the circulation. One could see that the dominant contribution in the small angle limit for the fermionic case is exactly the same as the bosonic one when the mass of the fermion is zero. This dominant contribution comes from the circulation term. Therefore, for small angles the contribution of the circulation is more important than the draing one. Furthermore, considering the cross section in the absence of the circulation shows that the result is the same for the bosonic and fermionic cases when the fermion is massless. In this limit, the result is independent of the angle $\phi$. 
The reason for that is in the absence of the circulation the only term that contributes in the long wavelength regime to the differential cross section is a polar symmetric term. Finally, we have shown that in contrast with the bosonic case in the limit $\omega \rightarrow 0$, considering $\alpha > \beta+\gamma^2$, the scattering cross section goes to a nonzero constant equal to $C \pi/2$ for the fermion case which is only dependent on the circulation.  

The similar scattering behavior for both boson and fermion for small angles seems to be in accord with both being scattered between the horizon and the ergosphere radius. At this regime both enjoys the Penrose effect of gaining energy after the scattering (the superresonance effect). As Figs.  \ref{fig1}-\ref{fig2} (right panel) show, this indeed agrees with the increasing of the differential cross section as the frequency increases. This precisely happens as long as the frequency belongs the interval $0<\omega<m\Omega_H$, where $m$ is  the azimuthal mode number and $\Omega_H$ is the angular velocity of the black hole \cite{ABP11,ABP2012-1}.  On the other hand, for large angles, both particles tend to be scattered outside the ergoregion and in turn they almost keep the scattering constant as Figs.  \ref{fig1}-\ref{fig2} (left panel) show. This is particularly more accurate for fermions. A detailed analysis of the superresonance effect for fermion fields should be addressed elsewhere. Besides that, in a future work, we plan to study a more realistic model considering the dynamics of the fermionic fields interacting with a black hole, gravitationaL or an analog in a condensed matter system. 

\section*{Acknowledgments}

We would like to thank CNPq and CAPES for partial financial support. A. M. thanks PNPD/CAPES for the financial support.

\end{document}